# Helix-like bio-polymers can act as dampers of force for bacteria in flows

Johan Zakrisson[a, b], Krister Wiklund[a, b], Ove Axner[a, b], Magnus Andersson[a, b,*]

[a]Department of Physics, Umeå University, [b]Umeå Centre for Microbial Research (UCMR), Umeå University, SE-901 87 Umeå, Sweden;

*Corresponding author: Magnus Andersson, Department of Physics, Umeå University,

SE-901 87 Umeå, Sweden, Tel: + 46 – 90 786 6336, FAX: +46 – 90 786 6673, E-mail: magnus.andersson@physics.umu.se

## ABSTRACT

Biopolymers are vital structures for many living organisms; for a variety of bacteria adhesion polymers play a crucial role for the initiation of colonization. Some bacteria express, on their surface, attachment organelles (pili) that comprise of subunits formed into stiff helix-like structures that possess unique biomechanical properties. These helix-like structures possess a high degree of flexibility that gives the biopolymers a unique extendibility. This has been considered beneficial for piliated bacteria adhering to host surfaces in the presence of a fluid flow. We show in this work that helix-like pili have the ability to act as efficient dampers of force that can, for a limited time, lower the load on the force-mediating adhesin-receptor bond on the tip of an individual pilus. The model presented is applied to bacteria adhering with a single pilus of either of the two most common types expressed by Uropathogenic *Escherichia coli*, P or type 1 pili, subjected to realistic flows. The results indicate that the force experienced by the adhesin-receptor interaction at the tip of the pilus can for moderate flows (~25 mm/s), be reduced by a factor of ~6 and ~4, respectively. The uncoiling ability provides a bacterium with a "go with the flow" possibility that acts as a damping. It is surmised that this can be an important factor for the initial part of the adhesion process, in particular in turbulent flows, and thereby be of use for bacteria in their strive to survive a natural defense such as fluid rinsing actions.

# INTRODUCTION

Biopolymers are large macromolecules produced by living organisms that are composed of repeating monomeric subunits. Although they have well-defined primary and second structures, they frequently and spontaneously arrange into characteristic higher order (tertiary and sometimes even quaternary) structures that often have a decisive impact upon their biomechanical properties. They play a variety of roles in biological systems. For example, those that are expressed on the outer membrane of bacteria and convey on their tip an adhesin that binds to specific receptors on the surface of host cells, alternatively referred to as attachment organelles, fimbria, adhesive fibers, or pili, are assumed to have an important role in the initial attachment of bacteria to host tissue (Duncan et al. 2005). To adapt to different environments, presumably because of evolutionary selection, various types of bacteria express organelles with different types of higher order structures. Whereas some are shaped like an open coil (linear chain) (Hilleringmann et al. 2008) others have a higher complexity with a helix-like quaternary shape (Bullitt and Makowski 1995). This gives various types of organelles dissimilar biomechanical properties (Axner et al. 2011; Castelain et al. 2009).

The biomechanical properties of attachment organelles are often assessed by their force-extension response by the use of force spectroscopic techniques, e.g., optical tweezers (Andersson et al. 2008; Chen et al. 2011) or atomic force microscopy (Lugmaier et al. 2007; Miller et al. 2006). Such studies have revealed that open coil-like polymers show a compliance that is dominated by elasticity and entropy, where the latter originates from random configurations of segments in the polymer giving rise to responses well described by the extended worm like chain model (WLC) (Bustamante et al. 1994). Extension of these types of polymers gives rise to responses similar to that of a non-linear spring, which for small elongations is characterized by a slowly increasing force whereas for long elongations, in the proximity of the polymers contour length, the force increases rapidly with extension.

In contrast, helix-like attachment organelles (pili) display a much richer wealth of response. The basic reason is that they consist of a repeated number of subunits that are arranged into a rigid helix-like arrangement, commonly referred to as the "rod" (Bullitt and Makowski 1995). This structure originates from a stabilization of subsequent turns of the polymer by weak turn-to-turn (TT) interactions formed by residues in each subunit (Bullitt and Makowski 1998). In general, there are multiple TT interactions per turn. This implies that a tensile force to which the pilus is exposed will, in the interior of the rod, be distributed among several interactions, which in turn implicates that each such interaction will experience solely a fraction of the force. The interaction that connects the outermost subunit of the helix-like structure, on the other hand, will experience a significantly higher force, close to the entire force. This gives rise to a sequential breaking of consecutive TT interactions starting with the outermost subunit, and thereby a sequential uncoiling of consecutive turns, that provides an extension that takes place under constant force (Jass et al. 2004). In addition, since the TT interactions connect subunits in a non-sequential manner, typically the *n*th subunit is connected to the (*n+3*)th unit, a helix-like pilus can be extended considerably, several times (5 – 10) its original length (Jass et al. 2004). This provides helix-like pili exposed to an external force with a unique biomechanical behavior; significant extension under infinite compliance (Axner et al. 2011). An example of such a response is given in Fig. 1 in the Supplementary material.

It has also been shown that the uncoiling process is reversible; an extended (uncoiled) structure will spontaneously recoil when the force decreases below the steady-state uncoiling force (Fällman et al. 2005). When the uncoiling and recoiling take place at slow extension speeds, or at forces close to the steady-state uncoiling and recoiling force, there is a balance between the opening and closure rates of the TT bonds. Under these so called steady-state conditions the uncoiling and recoiling processes are fully time-reversible (they take place at the same force). For higher extension speeds, the balance between the opening and closure rates of the bonds is no longer maintained, which leads to dissimilar uncoiling and recoiling forces (with the former being larger than the latter). This is referred to as a dynamic response (Andersson et al. 2006a). The relaxation rate, which assesses how fast a dynamically extended pilus retracts to its steady-state force when the forced extension has been halted (Andersson et al. 2008; Castelain et al. 2011), is an additional measure of the dynamic behavior of pili.

Various models for the biomechanical response of helix-like pili, all based on the energy landscape picture, e.g. the sticky-chain model or Monte Carlo simulations, have been presented in the literature. They have all proven to describe the observed behavior well (Andersson et al. 2006c; Andersson et al. 2007; Björnham et al. 2008). An example of a fit of a Monte Carlo simulation to a force-extension curve of helix-like pili is shown in Fig. 2 in the Supplementary material. Such models can provide information about the interactions involved, e.g. bond lengths and energies (Axner et al. 2010; Castelain et al. 2011), and serve as a basis for simulations of multi-pili responses (Björnham and Axner 2009; Björnham et al. 2008).

It has been hypothesized that the uncoiling and recoiling processes are beneficial for bacteria exposed to significant fluid forces (Andersson et al. 2008). They are considered to provide bacteria adhering to a host surface by multiple pili with an extraordinary ability to cooperatively redistribute an external force among several attachment organelles (Björnham and Axner 2009). Distributing the force among several pili lowers the load on individual adhesin-receptor bonds by which the bacteria adheres to the host, which, in turn, can increase the adhesion lifetime significantly. Uncoiling and recoiling are therefore assumed to play a significant role for the ability of bacteria to sustain considerable external forces from body fluids; in particular Uropathogenic *Escherichia coli* (UPEC, shown in Fig. 1A) exposed to urine flows (Björnham and Axner 2009).

Figure 1 here.

Despite the aforementioned types of response, the quaternary structure of the rod provides helix-like pili with additional biomechanical properties of which not all have yet been identified or characterized in the literature. One is that they can act as efficient *dampers* of force that can lower the load on an individual force-mediating adhesin-receptor bond. The present paper provides a description and an analysis of the principle behind this ability.

First, a simple description of the phenomenon is given. Although the findings presented are general and valid for any type of helix-like polymer stabilized by TT interactions, the phenomenon will be exemplified on two model systems; UPEC bacteria expressing either P or type 1 pili. For simplicity, we have assumed that the bacteria attach to a host with solely one pilus. As is discussed below, this is not a restriction that has any significant impact upon the phenomenon investigated; it merely simplifies the analysis. The results illustrate that, in the presence of a flow, and for a limited amount of time (during the uncoiling process), a helix-like pilus can act as a shock-absorber that momentarily can dampen a force, thus reducing the load on the adhesin-receptor bond by which it adheres to the host. It is shown that for the pili considered (and when exposed to large fluid flows) the reduction of force on the force-mediating adhesin-receptor bond can be substantial (> 50%).

## THEORY AND MODEL SYSTEM

### The force acting on a bacterium exposed to a flow

Consider an individual bacterium close to a surface, exposed to a liquid flow. For typical conditions, i.e. bacteria with dimensions in the µm range, and flow velocities in the mm/s range, which, for example, can take place in the laminar sub-layer close to a surface, the Reynolds number, *Re*, is << 1. This implies that we can assume the bacterium to be exposed to a laminar flow (Happel and Brenner 1983). If the bacterium is modeled as a spherical object with a radius

$r$, the force to which it is exposed can then be described by Stokes law, which can be expressed as

$$F(v_{\text{rel}}) = 6\pi\eta r v_{\text{rel}}, \tag{1}$$

where $\eta$ is the viscosity of the fluid and $v_{\text{rel}}$ is the relative velocity of the bacterium (the sphere) with respect to the flow, which, in turn, is given by the difference between the velocity of the fluid flow, $v_{\text{f}}$, and that of the bacterium, $v_{\text{b}}$, i.e. by $v_{\text{f}} - v_{\text{b}}$.

Hence, as long as the bacterium follows the flow, i.e. as long as $v_b = v_f$, it does not experience any force. On the other hand, if the bacterium is firmly attached to the surface (e.g. by a stiff organelle), whereby $v_b = 0$, it will experience a Stokes drag force given by

$$F(v_{\text{f}}) = 6\pi\eta r v_{\text{f}}. \tag{2}$$

This implies that if a free flowing bacterium expressing stiff organelles suddenly attaches to the surface by one of its organelles, the bacterium, and thereby the adhesion-receptor bond by which the organelle adheres to the surface, will instantaneously be exposed to the drag force given by Eq. (2).

However, if the organelle has a compliance (implying that it can be extended when exposed to a force) the bacterium would, as long as its organelle elongates, drift with the flow although with a reduced velocity, given by the rate of extension of the organelle (which thus can take a value $0 < v_b < v_f$). This implies, according to Eq. (1), that while the pilus extends, the bacterium (and thereby the adhesion-receptor bond by which the pili attaches to the host) will experience a reduced force. This shows that extendable pili can act as shock absorbers that efficiently dampens out rapid spikes in the flow, including those that arise when a free-floating bacterium suddenly attaches or when the flow is turbulent.

**Uncoiling and recoiling of a helix-like pili**

A helix-like structure composed of subunits sequentially connected by head-to-tail (HT) interactions and transversally attached by turn-to-turn (TT) interactions can provide compliance in two ways. First, it can extend in a manner similar to that of a regular spring by extending the TT bonds. Secondly, it can uncoil into an open coil like fiber by breaking the TT interactions. As was shown by Andersson *et al.* (Andersson et al. 2006a), a helix-like structure that uncoil (and recoil) sequentially, as illustrated in Fig. 1B, will have a net uncoiling and recoiling rate solely given by those of the outermost subunit. Therefore, the extension velocity $\dot{L}$ of a helix-like pili under tensile stress, $F$, can be expressed in terms of the difference between the uncoiling and the recoiling rates of a single TT interaction, $k_{AB}(F)$ and $k_{BA}(F)$, times the opening length $\Delta x_{AB}$ of the interaction, i.e. as

$$\dot{L} = [k_{AB}(F) - k_{BA}(F)]\Delta x_{AB}, \tag{3}$$

where A and B represent the closed and open states in the energy landscape describing pili dynamics. The uncoiling and recoiling rates can, according to Bell (Bell 1978), be expressed as

$$k_{AB}(F) = k_{AB}^{th} e^{F\Delta x_{AT}\beta}, \tag{4}$$

$$k_{BA}(F) = k_{AB}^{th} e^{(V_0 - F\Delta x_{TB})\beta}, \tag{5}$$

where $k_{AB}^{th}$ is the thermal opening rate of the TT interaction, $\Delta x_{AT}$ is the bond length (defined as the distance between the ground state and the transition state T), $\beta = 1/kT$, where $k$ the Boltzmann's constant and $T$ is the temperature, $V_0$ is the energy difference between the ground and open states of the TT interaction, and $\Delta x_{TB}$ is the distance between the transition state and the open state. This implies that the uncoiling speed $\dot{L}$ can be expressed in terms of the applied force $F$ as,

$$\dot{L} = \Delta x_{AB} k_{AB}^{th} \left( e^{F\Delta x_{AT}\beta} - e^{(V_0 - F\Delta x_{TB})\beta} \right). \tag{6}$$

When the pilus is partly extended, but not subjected to any change in length (i.e. $\dot{L} = 0$), the uncoiling and recoiling rates are equal, i.e. $k_{AB}(F) = k_{BA}(F)$. This takes place when the pilus is exposed to a force, referred to as the steady state uncoiling force, $F_{ss}$, given by

$$F_{ss} = \frac{V_0}{\Delta x_{AB}}, \tag{7}$$

[obtained by equalizing the Eqs (4) and (5)] where $\Delta x_{AB}$ is given by $\Delta x_{AT} + \Delta x_{TB}$. The pilus will extended (i.e. $\dot{L} > 0$) when the uncoiling rate becomes larger than the recoiling rate, which takes place as soon as the force is larger than this steady state uncoiling force. For extension velocities up to the so-called corner velocity, $\dot{L}^*$, (defined below), the extension takes place at a force that does not significantly exceed the steady state uncoiling force. When this takes place, the extension is said to take place under steady state conditions.

However, for extension velocities above the corner velocity, or when the pilus is exposed to forces significantly larger than the steady state uncoiling force, the opening rate becomes significantly larger than the closing rate, i.e. $k_{AB}(F) \gg k_{BA}(F)$. This implies that the second term in Eq. (6) can be neglected with respect to the first, whereby the extension velocity can be expressed as

$$\dot{L} = \Delta x_{AB} k_{AB}^{th} e^{F \Delta x_{AT} \beta}. \tag{8}$$

Under these conditions, the extension velocity is strongly (exponentially) dependent on the force, which, in turn, implies that the uncoiling force depends logarithmically on the extension speed. When this takes place, the extension is said to take place under dynamic conditions. The corner velocity, which is given by

$$\dot{L}^* = \Delta x_{AB} k_{AB}^{th} e^{V_0 \beta \Delta x_{AT}/\Delta x_{AB}}, \tag{9}$$

is, in fact, the extension velocity that separates the two regions (steady-state and dynamic) defined above.

Finally, by use of these equations [(6), (7), and (9)], it is possible to express the uncoiling velocity in terms of the previously defined entities, the corner velocity and the steady-state uncoiling force, as,

$$\dot{L} = \dot{L}^* e^{(F-F_{ss})\Delta x_{AT}\beta}[1 - e^{-(F-F_{ss})\Delta x_{AB}\beta}]. \qquad (10)$$

This is thus an expression for the extension velocity in the presence of the force to which the pilus is exposed in terms of well-defined model parameters that is valid under both steady-state and dynamic conditions.

**A bacterium expressing helix-like pili in flow**

Since the velocity of an attached bacterium $v_b$ is related to, but also constrained to, the uncoiling velocity of a pilus, i.e. $v_b = \dot{L}$, the force acting on the bacterium [according to Eq. (1)] will be related to the fluid flow through the extension properties of the pili rod. By combining the Eqs 1 and 10, and under the assumption that the directions of the flow and the extension are parallel, it is possible to derive an expression that relates the fluid flow to the force acting on a bacterium. Note that although it is not possible to derive an analytical expression of the force as function of flow, i.e. $F(v_f)$, as is needed in this work, it is possible to derive an expression for the reverse, i.e. the flow as a function of force, $v_f(F)$. Such an expression can be written as

$$v_f = \frac{F}{6\pi\eta r} + \dot{L}^* e^{(F-F_{ss})\Delta x_{AT}\beta}[1 - e^{-(F-F_{ss})\Delta x_{AB}\beta}]. \qquad (11)$$

The required dependence, $F(v_f)$, can be obtained either by plotting the flow as a function of force with reversed axes or by solving the expression numerically (i.e. for the force that corresponds to a given flow).

Due to its finite uncoiling length a helix-like pilus can only be extended during a given time $\tau$, referred to as the uncoiling time, given by,

$$L_0 = \int_0^\tau \dot{L}(t)dt = \dot{L}\tau, \tag{12}$$

where $L_0$ is the maximum length of an extended pilus, and the last expression is valid only under a constant uncoiling velocity. Since the velocity of the bacterium is equal to the uncoiling velocity of the pilus by which it is attached, i.e. $v_b = \dot{L}$, the Eqs 1 and 12 can be used to derive an expression of the maximum uncoiling time of a pilus as a function of the force that acts on a bacterium that reads

$$\tau = \frac{L_0}{\dot{L}} = \frac{L_0 6\pi\eta r}{v_f 6\pi\eta r - F}. \tag{13}$$

**Model systems: P and type 1 pili**

In order to exemplify the phenomenon of damping, and the general description of it presented above, two model systems based on two specific types of helix-like pili, P and type 1, were considered. These types of pili are found on bacteria giving rise to urinary tract infections (UTI); in fact, they are the most common types of pili expressed by UPEC bacteria that worldwide cause UTI. It has been found that P and type 1 pili have a qualitatively similar structure—they both have a helix-like quaternary structure (Hahn et al. 2002)—although they have quantitatively different behaviors—the dimensions of the individual subunits and the lengths and energies of the interactions are dissimilar. The backbone of the pili is, for both types of pili, composed of subunits that are linked in a head-to-tail manner via a donor strand exchange in which a subunit donates its amino terminal extension to complete the fold of its neighbor. The quaternary structure has a helix-like shape in which pili in one turn is connected to an adjacent turn via several non-covalent TT interactions (see Fig 1B).

The biomechanical properties of these pili have been scrutinized in some detail at a single organelle level using sensitive force measurement techniques (Andersson et al. 2006a, c; Chen et al. 2011; Fällman et al. 2005; Jass et al. 2004; Lugmaier et al. 2007; Miller et al. 2006). For example, the force-extension response of P pili has been assessed using both atomic force microscopy (AFM) and force measuring optical tweezers (FMOT) (Fällman et al. 2004; Lugmaier et al. 2007). Whereas the former technique has been used to demonstrate that the pili structure is strong and flexible—the hydrophobic interactions linking subunits by the HT interaction and intra-subunit interactions are strong and difficult to break (Lugmaier et al. 2007)—the latter has been used to characterize the force-extension response in some detail (Andersson et al. 2008).

It has been found that the TT interactions are weaker than the HT interactions, and can therefore more easily be compromised. As is shown in Fig. 1 in the Supplementary material, and as was alluded to above, FMOT measurements have showed that a pilus can be extended in three distinct regions allowing a significant increase in length (5 – 10 times). The first region is characterized by a linearly increasing force with an extension roughly similar to that of a Hookean spring. When the stretching force approaches ~28 pN a sudden transition to a second region, consisting of a characteristic constant force-extension response (a so called plateau) that is attributed to the above-mentioned sequential uncoiling of consecutive turns, takes place. After all turns in the helical structure have uncoiled (whereby the helical structure has been transformed into a linearized fiber), the force once again increases, exhibiting a force-extension response with a pseudo-Hookean behavior, referred to as region three. This behavior originates from fast stochastic conformational changes of the individual subunits that give rise to an entropic force (Andersson et al. 2006c). In addition, force measurements have also shown that the pili can recoil to their native structure in a reversible manner, without losing elasticity. Indeed, it has been shown that P pili do not show any sign of biomechanical changes or fatigue even though uncoiled 70 times (Andersson et al. 2006b).

Since UPEC bacteria expressing P or type 1 pili are, in the urinary tract, subjected to fluidic flows that give rise to forces that both can compromise the quaternary structure of their pili and agree with the conditions for damping modeled above, these two well-characterized types of pili were used as model systems for the illustration of the concept of pili damping.

## RESULTS

### Dynamic response of a helix-like pilus

As was alluded to above, both of the two types of pili considered, P and type 1, have previously been characterized in some detail with respect to their bio-mechanical properties on a single organelle level by the use of force-extension investigations (Andersson et al. 2007). To provide a complete description of the biomechanical properties of a helix-like pilus, i.e. to assess all relevant entities in the model described above, those studies were performed under steady-state as well as under dynamic conditions (Andersson et al. 2006a, c). The difference in force response gave rise to dissimilar sets of model parameters for P and type 1 pili, e.g., bond lengths and state energies (Andersson et al. 2008).

Using these sets of model parameters, first the uncoiling force responses (as a function of extension velocity) for P and type 1 pili were constructed according to the model presented above (Eq. 10). Figure 2 shows that for low velocities, i.e., in the steady-state regime, the uncoiling force is independent of extension velocity (28 and 30 pN for the two types of pili investigated, P and type 1 pili respectively), whereas for high velocities, the force has a logarithmic dependence. More importantly for this work is that the corner velocity, $\dot{L}^*$, which is defined as the velocity that separates the steady-state and dynamic response regimes (the velocities for which the two asymptotic responses intercept), is widely different for the two types of pili considered, it is for P and type 1 pili 400 and 6 nm/s, respectively. Hence, type 1 pili enter the dynamic regime for significantly lower extension velocities (~70 times) than what P pili do. This implies that type 1 more often than P pili extend under dynamic conditions. The difference in corner velocity of pili is so specific that it is even possible to use it as a fingerprint in force extension studies (Castelain et al. 2011).

Figure 2 here.

**Impact of uncoiling on the force experienced by a bacterium tethered with a helix-like pilus in a flow**

As was alluded to above, under the present assumptions, for which the force and extension are colinear, the extension velocity of the pilus rod is considered equal to the flow velocity of the bacterium. This allows us to illustrate the extension force, Eq. 10, and the Stokes drag force, Eq. 1, in the same plot, both as a function of the velocity of the bacterium. Figure 3 provides such a plot. The two non-linear curves, solid and dashed (red and blue), illustrate the dependence of force on the extension velocity, given by the velocity of the bacterium (relative to the surface), of P and type 1 pili whereas the thin linear curves (gray) represent the Stokes drag force for various fluid flow velocities (as marked, 1 – 8 mm/s), given by Eq. 1. The linear decline in force of the latter ones originates from the reduction in relative velocity due to the motion of bacterium.

Figure 3 here.

For a given flow velocity a particular type of pili will experience a force given by the intercept of a non-linear force-extension curve and the corresponding linear Stoke drag force curve. This force is thereby also the force the adhesin-receptor interaction on the tip of the pilus will experience. The three types of markers, open circles, filled circles, and open squares, represent the forces experienced by bacteria exposed to fluid flow velocities of 4, 5, and 6 mm/s, respectively. By plotting such intercept points as a function of fluid flow velocity, the force response in Fig. 4 can be obtained. For clarity, the three specific markers in Fig. 3 are also represented in Fig. 4. Alternatively, since the intercept points in Fig. 3 are those that simultaneously fulfill the Eqs (1) and (10), and those were the basis for Eq. (11), Fig. 4 can alternatively be obtained by numerically solving the implicit Eq. (11).

**Reduction in drag force for a bacterium adhering with an uncoiling helix-like pili**

To assess the influence of pili uncoiling on the force experienced by the adhesin-receptor interaction, Fig. 4 displays the force-*vs.*-flow response of a bacterium in three different cases. The dashed dotted black curve represents the situation for a bacterium that is attached by a stiff linker (represented by Eq. 2). The solid red and the dashed blue curves illustrate the response when the bacterium is attached by a single pilus for the two types of pili considered, P and type 1, respectively [represented by the Eq. (11)]. As expected, the force on a bacterium with a stiff linker increases linearly with flow whereas the ones expressing uncoilable helix-like pili experience a reduced force.

Figure 4 here.

At low fluid flow velocities, for which the helix-like pili do not uncoil (or only uncoil under close to steady-state conditions), the tensile force acting on the pilus is equal to (or similar to) the Stokes drag force for a bacterium with stiff linker. Under these conditions the uncoiling capabilities of the pilus do not play any major role in reducing the force exerted by the flow. However, for larger flows, the pilus can start to uncoil whereby the force decreases significantly; the adhesin-receptor interaction of P pili will experience a ~20% reduction for flows of ~5 mm/s whereas the same reduction takes place for type 1 pili at ~8 mm/s. The decrease in force is significant (>50 %) for P and type 1 pili for flows above 10 and 15 mm/s, respectively, as illustrated in Fig. 4 by black arrows.

However, this reduction of force will only take place under the time for which the pilus uncoils. To assess this time we numerically solved Eq. (13) for various flows. Figure 5 displays the uncoiling time of a ~2 µm long P (solid red curve) and type 1 (dashed blue curve) pilus as a function flow velocity (under the assumption that the pili can extend a factor of 5, to a total length of 10 µm). For flows below the critical velocity, $v_f^c$, for which the flow drag force of a stationary bacterium is equal to the steady-state uncoiling force, $F_{ss}$, (equal to 28 and 30 pN for the two types of pili), which are represented by the thin vertical red and blue lines, the uncoiling time becomes infinite. For velocities slightly above this, for which the uncoiling proceeds slowly, the uncoiling time becomes finite but huge. For higher flows, the uncoiling time decreases rapidly. However, due to the dissimilar corner velocities, the two types of pili experience significantly dissimilar uncoiling times. For example, for a flow of 2 mm/s P pili will uncoil in slightly less than 10 s, whereas for type 1 pili the uncoiling time is almost two orders of magnitude longer, i.e. around 1000 s. The uncoiling times of the two pili, P and type 1, are below one second for flow velocities up to 2.5 and 4.5 mm/s, respectively. The two black arrows in Fig. 4, which represent the conditions for a 50% reduction of force, correspond to uncoiling times of 1.8 and 1.3 ms, which are illustrated with filled black circles in the inset Fig. 5.

Figure 5 here.

# DISCUSSION

Assembling a helix-like biopolymer of a given length is more costly, in terms of energy, than producing a linear coil like polymer of the same length. Nothing in nature would, during millions of years of evolution, produce an energy costly structure without having a specific need for it. Helix-like pili have several unique biomechanical properties that therefore presumably play an important role in their natural habitat. They are commonly expressed by UPEC bacteria that reside in fluidic environments in which attachment to a host surface is crucial for surviving. Such bacteria are a frequent and isolated cause of urinary tract infections worldwide. In addition, the urinary tract is a niche with a hostile and dynamically changing environment with urine flow that can expose bacteria to high shear stress (Jin et al. 2010). In order for bacteria to stay attached and colonize the urinary tract under such hydrodynamic challenging conditions it is a necessary prerequisite that the adhesion process, both the initial and sustained adherence, is efficient. This implies, among other things, that pili need to possess biomechanical properties that are important for a bacterium in the adhesion process. Not only should a bacterium first manage to adhere to a host surface, it should also stay attached during intermittently severe environmental conditions; *in vivo* in situations that expose a bacterium to rapidly changing fluid forces that subjects its pili to significant amounts of stress. Since adhesion is the initial step of colonization, in the struggle of drug development and in the development of novel anti-bacterial surfaces for the food industry and hospitals, it is therefore of importance to characterize in detail the various properties of bacterial adhesion organelles with the aim to compromise their optimized ability to resist fluid forces (Klinth et al. 2012).

The presumably most important property of helix-like pili is their ability to extend significantly through uncoiling. This provides pili with an ability to redistribute an external shear force among several pili in such a way that no individual pilus gets any excess force, which is assumed to be beneficial for bacterial adhesion. It has also been found that pili can uncoil and recoil a large number of times; extensive single organelle studies have also showed that P pili can be extended/contracted up to >70 consecutive times without showing any fatigue (Andersson et al. 2006b).

In this work we have scrutinized yet another property of helix-like pili; *viz.* their ability to provide damping when exposed to rapidly changing tensile forces. This property originates from the ability of pili to reversibly and repeatedly uncoil and recoil, and thus to move with the flow, when exposed to force. Since the drag force experienced by the bacterium is caused by the relative velocity of the bacterium with respect to the flow, this will reduce the drag force and cause a damping. This property is rather unique and, to our knowledge, does not take place in any other surface expressed biopolymer However, the uncoiling property of pili resembles, from a macroscopic point of view, the extension of membrane tethers pulled out from leukocytes (Thomas 2008). Leukocytes are transported in the blood flow and need transient interruptions in order to perform rolling exploration of vessel walls during inflammatory response (McEver 2002). It is therefore believed that tether formation will reduce the force acting on the interaction between P-selectin, expressed on endothelial cells or platelets, and the leukocyte mucin P-selectin glycoprotein ligand 1 (King et al. 2005). Uncoiling of pili is, in a similar way, assumed to be an important property for bacteria living in a hostile environment where, e.g., the rinsing action of urine is a natural defense mechanism, in particular since such flow can fluctuate significantly in strength. Interestingly, it has also recently been shown that enterotoxigenic *E. coli* expressing CFA/I pili, which is a common cause of diarrhea and found in the intestine where the churning motion creates vortices that moves the fluid back and forth, have similar biomechanical properties as those expressed by UPEC (Andersson et al. 2012).

The description of damping of pili given in this work is based upon previously developed biomechanical models of helix-like pili expressed by UPEC bacteria that reside in fluidic environments in which attachment to a host surface is crucial for surviving. We began to explore, using previous experimental data and theory, the force-*vs.*-extension-speed response of helix-like pili. We also introduced the corner velocity, $\dot{L}^*$, which separates the steady-state and the dynamic regime, since this was found to be of importance for the time under which a pilus can function as a damper. The impact of the uncoiling of a single helix-like pilus expressed by a bacterium in a flow, and in particular its dynamic response, was simulated with a model that approximated the bacterium as a sphere in a local uniform flow exposed to a Stokes drag force. Even though this is a simplification it does not significantly affect the major conclusions of this work regarding the influence of helix-like surface organelles on the force experienced by a bacterium in a flow.

Urine in humans are transported from the kidney to the bladder in boluses by a peristaltic activity and later expelled from the bladder via the urethra. These two types of flows have high mean velocities and rapid changes in their velocity profile. For example, the bolus causes a high variation of shear rates and mean flow velocities up to 30 mm/s (Kinn 1996) whereas the flow in the urethra can be turbulent and locally reach 1000 mm/s (Abrams 1997). Even though the mean velocities can be high the velocity near the surface will be significantly smaller due to wall friction. A surface attached bacterium exposed to such flows would thereby experience a rather low drag force, often insufficient to cause uncoiling of its pili. However, it has been shown that the peristaltic contractions in the ureter that cause a bolus induce a flow field that is complex with strong backflows and turbulent motion (Jeffrey et al. 2003; Vahidi et al. 2011). In addition, based on a calculation from *in vivo* data in the urethra (Abrams 1997) the Reynolds number in that region can be as high as 5000, which is above that for which a flow is considered to be fully turbulent (4000). For turbulent flows, like in boluses, and during expel of urine through the urethra, it is also commonly known that short bursts and sweeps can inject fluid from high velocity regions into the laminar layer and that the timescale for these bursts and sweeps are in the millisecond regime (Xu et al. 1996). Thus, in both the ureter and the urethra the turbulent flows can generate short bursts and sweeps that reach down close to the surface where the bacteria are attached (Smith et al. 1991). The fast moving fluid gives rise to fluid "tsunamis" that for short periods of time expose bacteria to significantly higher shear forces, which could cause bacteria to detach. It is interesting to note that the millisecond timescales of these are in line with the timescale for uncoiling of a pilus found in this work. Expressing a structure with an ability to uncoil during either the passage of a bolus in the ureter, which exposes the bacteria to a rapid changing fluid environment due to the vortex circulation, or in the turbulent flow in the urethra, can thus be a beneficial property that can help a bacterium to stay attached to a host cell.

As an example, a 2 µm large (diameter) bacterium attached with a single pilus of either P or type 1 type exposed to various flows was considered. It was found that for flows lower than the typical mean flow velocity in a bolus, *viz.* 10 and 15 mm/s for the two types of pili respectively, the force to which the adhesin-receptor interaction of the tip of the pilus is exposed is reduced by 50%. This is indicated by the black arrows in Fig. 4. For higher velocities, the reduction in force becomes larger; e.g. for a flow velocity of 25 mm/s, the reduction in force becomes > 50%; whereas a stiff linker would experience ~440 pN, the two types of pili (P and type 1) will experience ~75 pN and ~130 pN, respectively. This thus shows that the possibility to uncoil indeed can reduce significantly the drag force to which a pilus, and thereby an adhesin-receptor interaction, is exposed. It is also remarkable that the drag force on a bacterium equipped with a type 1 pilus levels out slightly above 100 pN for large fluid velocities. This value is similar to what has been found for the force that provides the maximum adhesion lifetime of the FimH catch bond (Aprikian et al. 2011).

In comparison to tether formation by leukocytes, which is believed to reduce the load on the adhering bond, it is notable that the transient interruption from rapid movement to adhesion takes place in < 0.01 s (King et al. 2005). Our work shows that pili uncoiling are effective dampers on similar time scales and the results indicate that pili have properties and function similar to those of tethers formed by leukocytes.

The consequence of events like bursts and sweeps on helix-like pili expressed on bacteria is a complex phenomenon that requires significantly experimental as well as theoretical investigations to be described and analyzed in detail. We have in this work focused on the general concept of damping and demonstrated that this dynamic effect indeed can take place in helix-like pili, and estimated under which conditions it can appear for bacteria attaching with a single pilus. A more complete investigation would either require a more detailed description of bursts and sweeps in the urine and how these would affect a bacterium, or a consideration of multi-pili attachment. Although it can be concluded from the analysis above that each adhesin-receptor interaction on a multi-pili attached bacterium would be exposed to a lower force and experience a longer uncoiling time than in the single-pilus attachment case, these matters are, however, outside the objectives of the present work and will be the subject of a future study.

## CONCLUSION

Both the initial attachment and the sustained adhesion of bacteria are vital processes for initiation of infections. In order to overcome natural defense mechanisms in a host, such as fluid flows, bacteria have develop surface organelles that both express the adhesin distal of the cell and have biomechanical properties that help reducing shear forces in an efficient way. It has previously been shown that the extraordinary property of pili allow an external force to be redistributed to a multitude of pili so that none will experience any excessive force. Here we have shown that the dynamic effect of pili, by their ability to act like dampers, also allows a single pilus to reduce the force load on the adhesin. We have specifically for P and type 1 pili shown that they work as efficient shock absorber for sudden changes or fluctuations in fluid flow velocities. Helix-like pili are thus powerful unique surface organelles that both are efficient for the initial attachment to a host and for sustained adhesion in a hostile environment such as the urinary tract.

We have thus in this work showed, by combining previous assessed experimental data together with hydrodynamic and rate theory, that the helix-like structure of pili can provide a bacterium with important advantages for the initial attachment to host cells and efficiently support sustained adhesion in turbulent fluid flows.

## ACKNOWLEDGEMENT

This work was performed within the Umeå Centre for Microbial Research (UCMR) Linnaeus Program supported from Umeå University and the Swedish Research Council (349-2007-8673). It was also supported by a Young Researcher Award (*swe.* Karriärbidrag) from Umeå University (to M.A.) and by the Swedish Research Council (to O.A, 621-2008-3280). We are grateful to Bernt Eric Uhlin's research group for providing the AFM image of the *E. coli*.

# FIGURE CAPTIONS

**Figure 1.** A) AFM micrograph of an uropathogenic *E. coli* (HB101/pPap5) cell expressing P pili. These pili are 6–7 nm in diameter and ~2 µm long helix-like rods composed of a repeating number of subunits (PapA) ~1000 in a right-handed helical arrangement with 3.28 subunits per turn. B) Schematic illustration of the structural subunits attached HT in a pilus seen from a side and top view. The left most part resides in a helix-like configuration (visually enhanced in top view), with turn-to-turn interactions between the *n*th and (*n+3*)th subunits, whereas the right part illustrates the uncoiled configuration. Uncoiling under tensile stress is sequential due to the lower probability of opening an interaction inside the helical rod than opening the outermost interaction in the rod.

**Figure 2.** The force response of a P (solid red) and type 1 (dashed blue) pilus as a function of extension velocity. Data was obtained by solving Eq. 6. The intersections (dots) of the dashed lines represent the corner velocities, which are 400 and 6 nm/s for P and type 1 pili, respectively. Thus, the corner velocity divides the velocity dependent uncoiling force into a steady-state and dynamic regime.

**Figure 3.** The force response of a bacterium expressing P (solid red) and type 1 (dashed blue) pilus as a function of bacterium velocity. The two non-linear curves, solid and dashed (red and blue), represent the force-extension response of P and type 1 pili respectively, whereas the thin linear curves (gray) represent Stokes drag force for various fluid flow velocities (as marked, 1 – 8 mm/s), given by Eq. (1). Since $v_{\text{rel}}$ decreases with increasing uncoiling velocity $\dot{L}$, and thereby the bacterial velocity, $v_b$, the force experienced by a bacterium will thus decrease when a pilus uncoils. The three types of markers, open circles, filled circles, and open squares,

represent the forces experienced by bacteria exposed to fluid flow velocities of 4, 5, and 6 mm/s. As an example, the grey line denoted with a 5 represents a fluid flow of 5 mm/s that gives rise to a drag force of 88 pN to a bacterium attached with a stiff linker. In addition, the respective force acting on a bacterium with pili that can uncoil is at this flow decreased significantly (from 88 to 68 pN) for P pili and marginally (from 88 to 87 pN) for type 1. All velocities are relative to the surface, and thereby the adhesive tip.

**Figure 4.** The drag force as a function of flow for a bacterium attached to a surface with a stiff linker (dashed dotted black line), a P (solid red) and a type-1 pilus (dashed blue). The three markers represent the same particular situations as marked in Fig. 3. For low flows (<2 mm/s) the fluid force acting on a bacterium is insufficient to uncoil a pilus, whereby a pilus behaves similar to that of a stiff linker. However, when the flow is increased the tensile stress reaches the steady-state uncoiling force and the pilus starts to uncoil giving rise to the smooth decrease in force. The decrease in force is significant (>50 %) for P and type 1 pili for flows above 10 and 15 mm/s as illustrated in by the black arrows. For a high flow of 25 mm/s the force is reduced from 440 pN for a stiff linker to ~75 pN with a P and ~130 pN for a type 1 pilus.

**Figure 5.** A simulation of the total time needed to uncoil a P (solid red) and type 1 (dashed blue) pilus from 2 to 10 µm for a given flow. For the smallest flows required to uncoil a pilus (close to its steady-state force, in this case <2 mm/s) the uncoiling time is long, marked by the vertical dashed lines. For larger flows, the uncoiling time decreases rapidly. For the flows for which the reduction of force is >50% (marked with filled circles in the inset), the uncoiling times for the two types of pili, P and type 1, are 1.8 and 1.3 ms, respectively.

# FIGURES

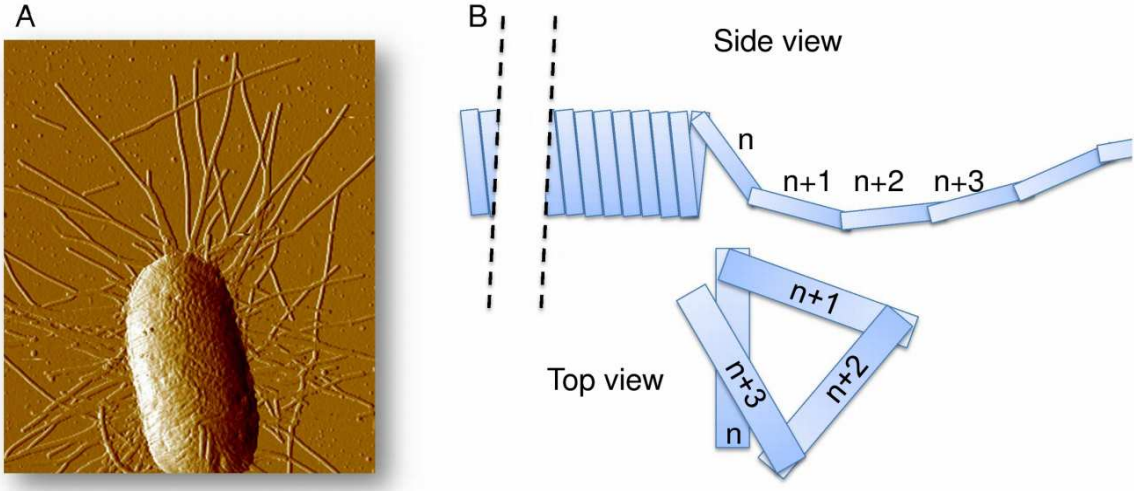

**Figure 1.**

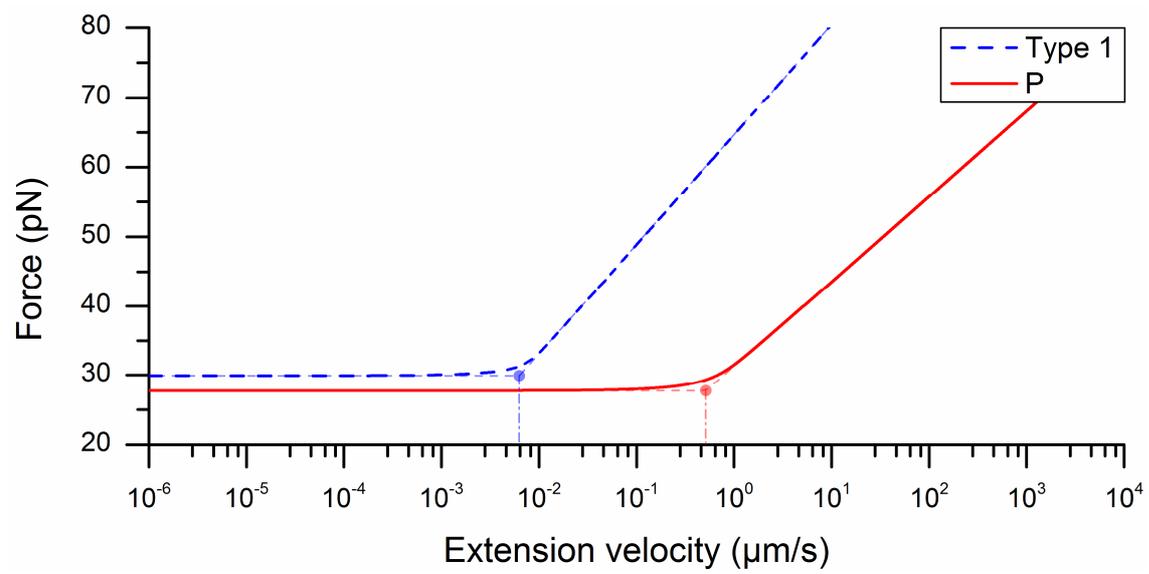

**Figure 2.**

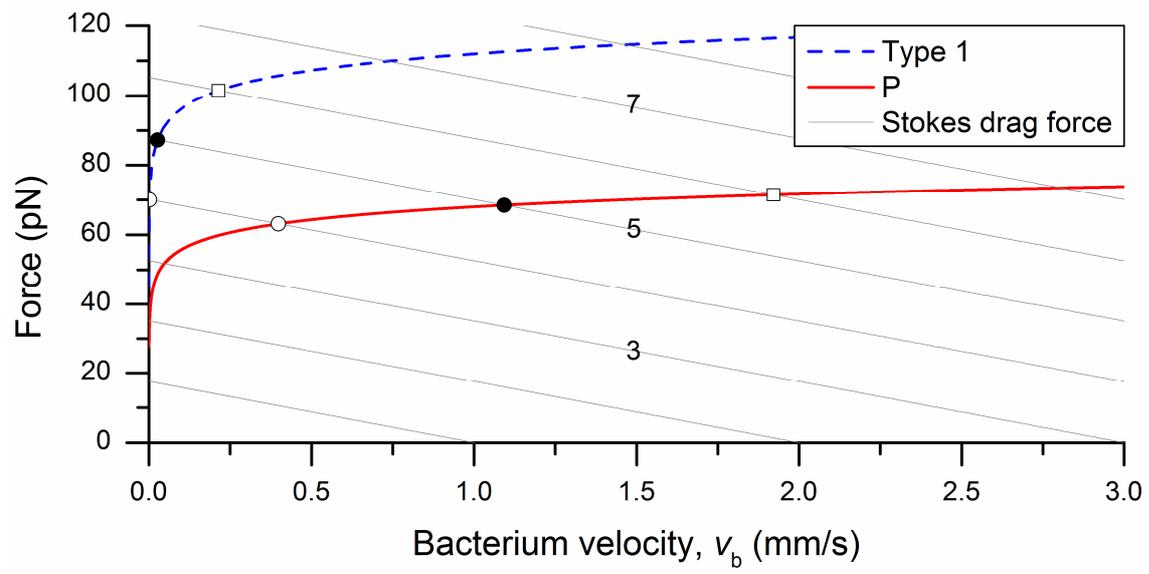

**Figure 3.**

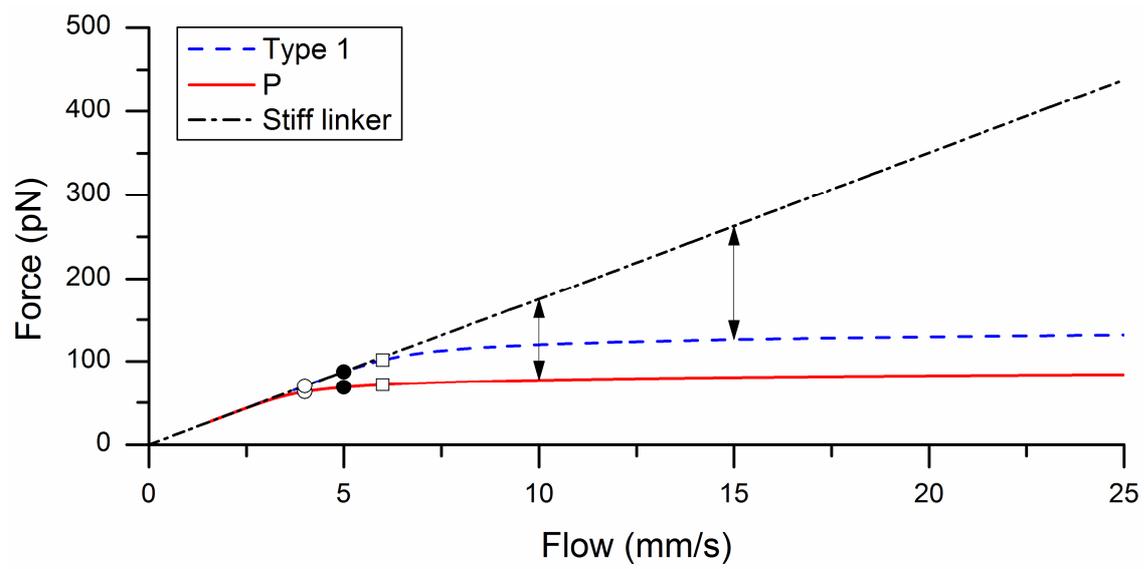

**Figure 4.**

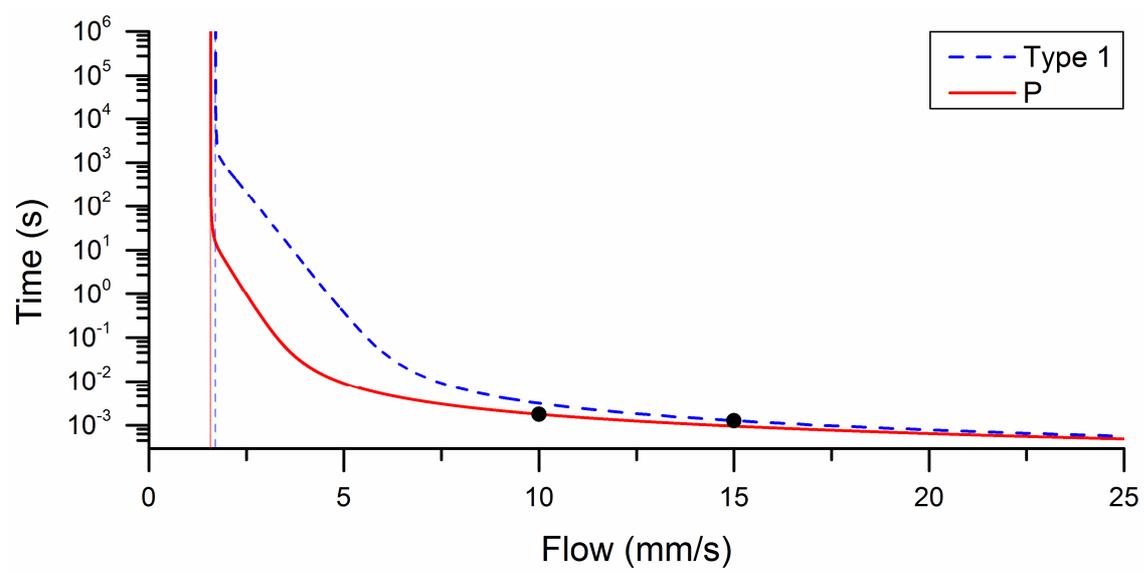

**Figure 5.**